\documentclass{PoS}

\title{Electromagnetic follow-up of gravitational wave candidates: perspectives in INAF - Italy}

\ShortTitle{Electromagnetic follow-up of gravitational wave candidates: perspectives in INAF - Italy}

\author{S. Piranomonte$^{a}$, E. Brocato$^{a}$, M. Branchesi$^{b,c}$, S. Campana$^{d}$, E. Cappellaro$^{e}$, S. Covino$^{d}$, A. Grado$^{f}$, E. Palazzi$^{g}$, L. Nicastro$^{g}$, E. Pian$^{h}$, G. Stratta$^{b,c}$, G. Greco$^{b,c}$, M. Castellano$^{a}$, G. Giuffrida$^{i,a}$, S. Marinoni$^{i,a}$, L. Pulone$^{a}$, A. Antonelli$^{a}$, M. G. Bernardini$^{d}$, P. D'avanzo$^{d}$, A. Melandri$^{d}$,  and L. Stella$^{a}$ on behalf of a larger collaboration\\
     \llap{$^a$}INAF - Osservatorio Astronomico di Roma, Monte Porzio Catone (RM), Italy\\
     \llap{$^b$}Università degli studi di Urbino, Urbino, Italy\\
     \llap{$^c$} INFN, Sezione di Firenze, I-50019 Sesto Fiorentino, Firenze, Italy\\
     \llap{$^d$}INAF - Osservatorio Astronomico di Brera, Milan, Italy\\
     \llap{$^e$}INAF - Osservatorio Astronomico di Padova, Padova, Italy\\
     \llap{$^f$}INAF - Osservatorio Astronomico di Capodimonte, Napoli, Italy\\
     \llap{$^g$} INAF-Istituto di Astrofisica Spaziale e Fisica Cosmica, Bologna, Italy \\
     \llap{$^h$}Scuola Normale Superiore, Pisa, Italy \\
     \llap{$^i$}ASI/ASI Science Data Center, Rome, Italy\\
     E-mail:  \email{silvia.piranomonte@oa-roma.inaf.it}}

\abstract{The electromagnetic (EM) emission associated with a gravitational wave (GW) signal is one of the main goal of future astronomy. Merger of neutron stars and/or black holes and core-collapse of massive stars are expected to cause rapid transient electromagnetic signals. The EM follow-up of GW signals will have to deal with large position uncertainties. The gravitational sky localization is expected to be tens to hundreds of square degrees. 
Wide-field cameras and rapid follow-up observations will be crucial to characterize the EM candidates for the first EM counterpart identification.
We present some of the activities that we are currently carrying on to optimize the response of the INAF network of facilities to expected GW triggers. The INAF network will represent an efficient operational framework capable of fast reaction on large error box triggers and direct identification and characterization of the candidates.}

\FullConference{Swift: 10 Years of Discovery,\\
		2-5 December 2014\\
		La Sapienza University, Rome, Italy }

\begin{document}

\section{Introduction}
The long-standing quest for gravitational waves (GW) detection from high energy celestial sources, cosmic explosions and astrophysical transients in general may soon meet with success when, within the next decade, a worldwide network of advanced versions of ground-based GW interferometers will become operational within the frequency range of 10 Hz to a few kHz \cite{Harry2010, Acernese2008, Accadia2012, Aasi2013}. The joint Advanced Laser Interferometer Gravitational Wave Observatory (LIGO) Scientific Collaboration and the Virgo Collaboration (LVC) will start regular and systematic operations of their sensitive, upgraded interferometers in 2015. Besides representing the single still unaccomplished direct test of general relativity, the detection of GWs allows insight into radiation processes, explosion mechanisms, stellar structure in a way that is independent from intrinsic opacity issues, extrinsic absorption and propagation effects that often plague observations of electromagnetic type. GWs thus probe the inner regions of many astrophysical phenomena that are otherwise inaccessible to investigation. Thus, the detection of GWs and their electromagnetic counterparts will definitively sign the birth of the Gravitational Astronomy and the availability of a new powerful messenger that, together with the two other messengers, photons and neutrinos,  will bring us unexplored evidences on the Physics of the matter in the Universe.

\section{GW sources detectable by LIGO/Virgo}
Most promising GW emitters appear to be coalescing neutron-stars (NS) and/or black-holes (BH) binary systems \cite{Peters1964, Abramovici1994} because the rapid rotation and large mass densities produce characteristic signal levels expected to be detectable by Advanced LIGO-VIRGO, if occurring in the local Universe. Several observation evidences support the hypothesis that compact binary star mergers are at the origin of short gamma-ray bursts (SGRBs).  
Other candidates for GW sources in the range of frequencies accessible to ground-based GW interferometers are also long GRBs \cite{corsimezsaros2009}, core-collapse supernovae \cite{Ott2009, Ott2012}, newly born magnetars \cite{stella2005, dallosso2009} and magnetar flares \cite{corsiowen2011, abadie2011}. However, the GW emission and the rates of GW detection predicted for these sources are highly model-dependent and therefore highly uncertain; moreover the expected signal levels are rather low and, for some source classes, only detectable for Galactic events.
LIGO and VIRGO detectors are currently being upgraded to their advanced configurations \cite{Harry2010, Acernese2008, Accadia2012} resulting in a sensitivity improvement of a factor ten and a corresponding increase in the number of detectable sources of a factor of 1000. The rate for NS binary coalescence with the advanced GW detectors is expected to be 40 per year \cite{Aasi2013, Abadie2012}.
The rate of massive stars core-collapse is of the order of 100-1000 events per year within a distance of 100 Mpc \cite{guettadellavalle2007}. However, the GW detectable rate of core-collapse is unknown mainly due to the uncertainties on the amount of energy emitted in GWs.


\begin{figure}
\centering
     \includegraphics[width=.6\textwidth]{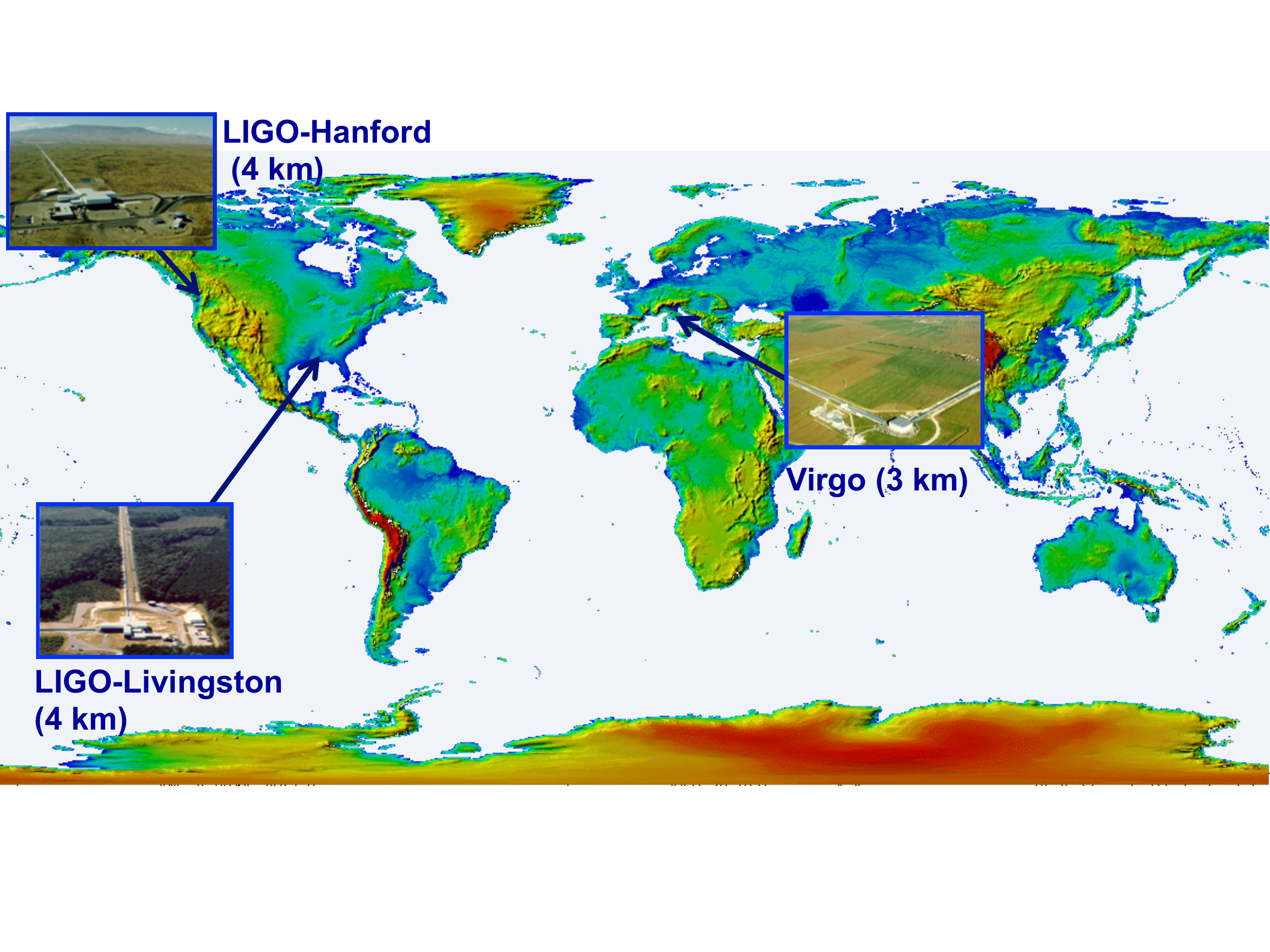}
     \caption{Location of the advanced ground-based gravitational wave detectors, which will observe the sky as a single detector.}
     \label{fig1}
     \end{figure}

\section{The international scenario}
The astronomical community is getting organized for the search and follow-up of GW sources with ground-based radio, optical, infrared and atmospheric Cherenkov telescopes and from space. The Italian National Institute of Astrophysics (INAF) has signed the MoU with LVC providing to the italian astronomical community the advantage of participating to this effort with the support of a whole institution. The observing forces that Italy will bring to bear include national telescopes controlled by INAF (e.g. CITE, TNG, REM, SRT, ASTRI, etc.) and large telescopes participated by INAF through consortia (ESO telescopes, among which major player are the VST and the VLT, LBT, CTA).

\section{General Perspectives}
The immediate goal is to organize and optimize the response to the triggers of GW signals provided by LVC. Among the points of strength within INAF there are: i) the worldwide recognized expertise and know-how on the characterization of transient and on high quality photometry in CCD images and ii) the existing international collaborations developed by GRB and time domain astronomy INAF communities. 
On the operative side specific efforts has to be performed to set up an observational strategy and efficient operational framework taking into account the different telescope capabilities. For example, the EM candidates identified by the wide-field imagers in Europe could be characterized by smaller facilities in Italy, by the Canary Island telescopes and by LBT. In the South hemisphere REM and TORTORA could work in coordination with the VST telescope: TORTORA is able to cover the GW error box at brighter magnitudes and VST could start the observations from sky regions with higher probability to contain the GW sources. REM can ensure simultaneous optical and infrared observations of the EM counterpart candidates. Clearly, a key role will be played by VLT for the identification of the unique EM counterpart. In space the observational strategies will be driven by the deep expertise of the Italian astronomers on the X-ray counterpart detection of the GRBs, preparation of Target of Opportunity (ToO) proposals to ground (e.g. ESO) and space observational facilities which have been already started thanks to previous experience and contacts of the GRB community.

\begin{figure}
\centering
     \includegraphics[width=.6\textwidth]{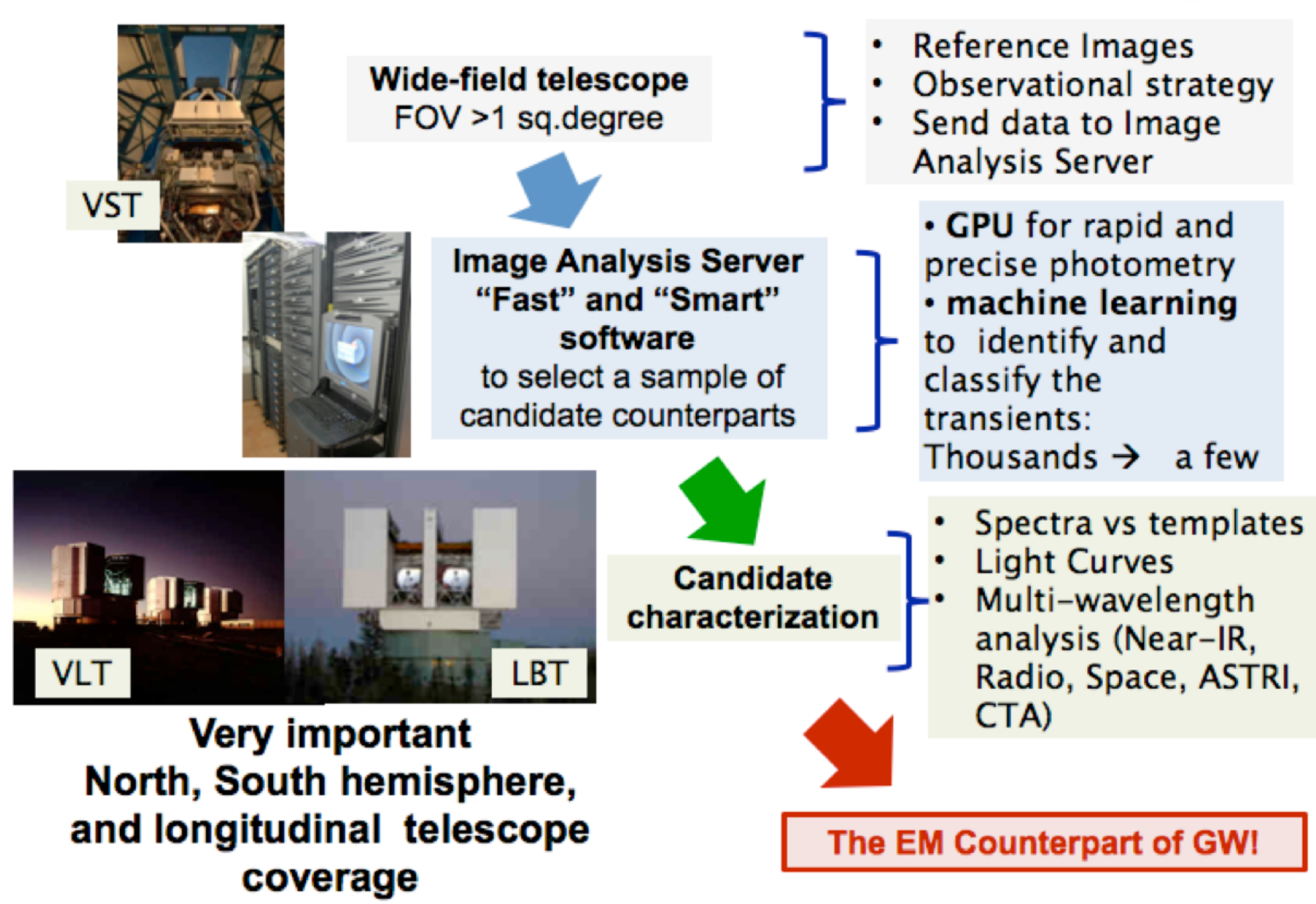}
     \caption{Required steps for an efficient EM-follow up of gravitational wave signal.}
     \label{fig2}
     \end{figure}

The organization, coordination and optimization of ground-based facilities requires intense and sophisticated activities (e.g. changes/updates/upgrades for the telescopes/instruments optimization, observational strategy set-up, dedicated software development for fast data analysis, data archive implementation and management, etc.), which the Italian community can offer due to the unique expertise previously matured in nearly 20 years of multi-wavelength (from radio to gamma-ray) follow-up of GRB sources.
Once a precise (arcsecond) localization of objects like GRBs is given, it will be possible to identify the host galaxy and have a multi-messenger picture for a complete knowledge of the most energetic events in the Universe.
In general GW and EM together will provide insight into the physics of the progenitors (mass, spin, distance etc.) and their environment (temperature, density, redshift etc.). 

\section{Conclusions}
The new technological steps and collaboration opportunities (MoU) developed by LIGO/Virgo Collaboration are opening new promising and concrete challenges in the interdisciplinary field of the GW astronomy.  The identification of EM counterparts of GW signal is clearly an issue which is extremely demanding efforts for astronomical observations. For this reason, it will be important to enhance the interdisciplinary exchange of know-how and the european collaboration between the worldwide community related to the multi-messenger astronomy.  The efforts in improving the data analysis techniques and in pushing the instrumentation at their extreme limits to cover very large field of view with digital CCD images is expected to trigger development in all the fields related to the Time Domain Astrophysics (TDA) and sciences that make use of the image recognition techniques. 
Finally, all the activities described above are expect to provide means and opportunities to the Italian and European astronomical communities to have a leading role in the GW astronomy and TDA fields.

\end{document}